# Fluid Flow along the Riga Plate with the Influence of Magnetic Force in a Rotating System

Muhammad Minarul Islam[1, a)], Sheela Khatun [1, b)], Md. Tusher Mollah [1, c)] and Md. Mahmud Alam[2, d)]

[1] *Department of Mathematics, Bangabandhu Sheikh Mujibur Rahman Science and Technology University Gopalganj-8100, Bangladesh*
[2] *Mathematics Discipline, Khulna University, Khulna-9208, Bangladesh*

a) Corresponding author: minarul_math@yahoo.com
b) sheela.bsmrstu@gmail.com
c) tusher.bsmrstu@gmail.com
d) alam_mahmud2000@yahoo.com

**Abstract.** The fluid flow along the Riga plate with the influence of magnetic force in a rotating system has been investigated numerically. The governing equations have been derived from Navier-Stokes' equations. Applying the boundary layer approximation, the appropriate boundary layer equations have been obtained. By using a usual transformation, the obtained governing equations have been transformed into a coupled dimensionless non-linear partial differential equation. The obtained dimensionless equations have been solved numerically by an explicit finite difference scheme. The simulated results have been obtained by using MATLAB R2015a. Also, the stability and convergence criteria have been analyzed. The effect of several parameters on the primary velocity, secondary velocity, temperature distributions as well as the local shear stress and the Nusselt number have been shown graphically.

## INTRODUCTION

Riga plate is known as the electromagnetic actuator which is the combination of permanent magnets and a spanwise aligned array of alternating electrodes mounted on a plane surface. It can be used for the radiation of an efficient agent; skin friction and pressure drag of submarines by avoiding the boundary layer separation. Flows under the influence of magnetic force along the Riga plate play a fundamental role in various industrial and engineering processes, such as MHD generators, thermal nuclear reactors, flow meters and the design of nuclear reactors. It is well known that such flows have tremendous applications in civil engineering, mechanical engineering, chemical engineering, food processing and in biomechanics. In this regard, the characteristics of laminar fluid flow due to Riga plate has been investigated in various physical aspects. Gallites and Lilausis [1] formulated a Riga plate to build an applied magnetic and electric field which consequently generates a Lorentz force parallel to the wall in order to control the flow of fluid. The behavior of fluid flow having low electrical conductivity has been investigated by Pantokratoras and Magyari [2]. Abbas, et al. [3] investigated the entropy generation on nanofluid flow through a horizontal Riga plate. The flow of nanofluid due to convectively heated Riga plate with variable thickness has been studied by Hayat, et al. [4]. Iqbal, et al. [5] considered the melting heat transport of nanofluidic problem over a Riga plate with erratic thickness. The non-linear Radiative flow of nanofluid past a moving or stationary Riga plate has been studied by Ramesh and Gireesha [6]. Anjum, et al. [7] considered the influence of thermal stratification and slip conditions on stagnation point flow towards variable thicked Riga plate.

Hence the aim is to numerically investigate the unsteady viscous incompressible fluid flow along the Riga plate with the influence of magnetic force in a rotating system. The usual transformations have been used to obtain the non-dimensional coupled non-linear partial differential equations. The explicit finite difference technique has been used to solve the dimensionless governing equations. The obtained results have been shown graphically.

# MATHEMATICAL FORMULATION

An unsteady and viscous incompressible fluid flow with the presence of magnetic force is considered for a rotating system. The flow is generated by a Riga plate situated at $y=0$ and having the $X$-axis vertically above as shown in Fig. 1. The flow is driven over the horizontal plate only by a Lorentz force $\mathbf{f} = \sigma(\mathbf{E} \wedge \mathbf{B})$, where $\sigma$ is an electrical conductivity of the fluid, $\mathbf{B}$ is the magnetic induction, and $E$ is the electric field directed parallel to the array. The wall temperature is $T_w$ and the temperature outside the boundary layer is $T_\infty$; where, $T_w > T_\infty$. In the case of the Riga-plate, the edge-effects are neglected, both the applied electric and magnetic fields have components only in the wall-normal direction $Y$ and the spanwise direction $z$. Thus the fluid velocity vector is, $\mathbf{q} = u\hat{i} + v\hat{j} + w\hat{k}$.

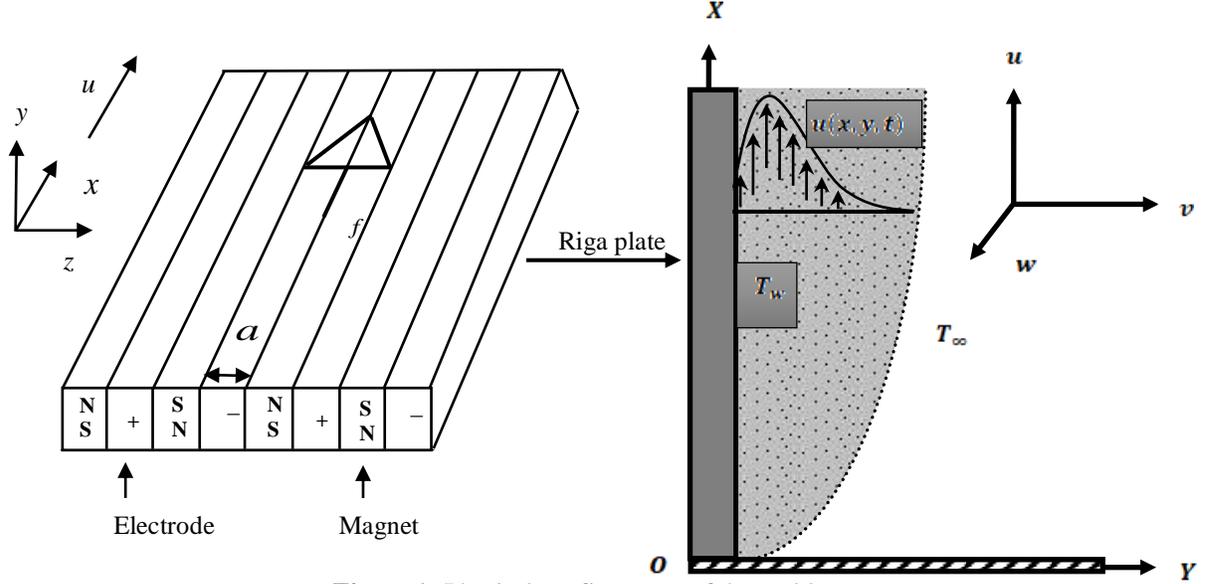

**Figure 1:** Physical configuration of the problem

To obtain the dimensionless governing equations, the following non-dimensional variables are taken as follows:

$$X = \frac{x}{l}, Y = \frac{y}{L}, U = \frac{u}{U_0}, V = \frac{v}{V_0}, W = \frac{w}{U_0}, \tau = \frac{tU_0}{l}, \theta = \frac{T - T_w}{T_\infty - T_w}; \text{ where, } V_0 = \frac{\upsilon\pi}{a}, l = \frac{U_0 L^2}{\upsilon}, L = \frac{a}{\pi}$$

Using these above dimensionless variables, the dimensionless governing equations have been obtained as follows:

$$\frac{\partial U}{\partial X} + \frac{\partial V}{\partial Y} = 0 \qquad (1)$$

$$\frac{\partial U}{\partial \tau} + U\frac{\partial U}{\partial X} + V\frac{\partial U}{\partial Y} = \frac{\partial^2 U}{\partial Y^2} + Ze^{-Y} + 2RW \qquad (2)$$

$$\frac{\partial W}{\partial \tau} + U\frac{\partial W}{\partial X} + V\frac{\partial W}{\partial Y} = \frac{\partial^2 W}{\partial Y^2} + Ze^{-Y} - 2RU \qquad (3)$$

$$\frac{\partial \theta}{\partial \tau} + U\frac{\partial \theta}{\partial X} + V\frac{\partial \theta}{\partial Y} = \left(\frac{1}{P_r} + \frac{4}{3}R_d\right)\frac{\partial^2 \theta}{\partial Y^2} + E_c\left[\left(\frac{\partial U}{\partial Y}\right)^2 + \left(\frac{\partial W}{\partial Y}\right)^2\right] \qquad (4)$$

The corresponding non-dimensional boundary conditions can be written as follows:
$t > 0$, $U = 0, W = 0, \theta = 0$ at $Y=0$ and $U \to 0, W \to 0, \theta \to 1$ at $Y \to \infty$

Where the non-dimensional parameters are given as follows:

Rotational parameter, $R = \frac{\Omega L^2}{\upsilon}$; Modified Hartmann number, $Z = \frac{J_0 M_0 a^2}{8\pi\rho U_0 \upsilon}$; Prandtl number, $P_r = \frac{\upsilon \rho c_p}{\kappa}$; Eckert number, $E_c = \frac{\mu l U_0}{\rho c_p L^2 (T_\infty - T_w)}$; and Radiation parameter $R_d = \frac{4\sigma^* T_\infty^3}{k^* \rho c_p}$.

## SHEAR STRESS AND NUSSELT NUMBER

From the velocity field, the effects of various parameters on the shear stresses have been calculated. The following equations represent the local shear stresses at the plate. For primary velocity, the local shear stress in $X$-direction i.e. the local primary shear stress is, $\tau_{LX} = \mu\left(\frac{\partial U}{\partial Y}\right)_{Y=0}$. For secondary velocity, the local shear stress in $Z$-direction i.e. the local secondary shear stress is, $\tau_{LZ} = \mu\left(\frac{\partial W}{\partial Y}\right)_{Y=0}$. From the temperature field, the effects of various parameters on the Nusselt number have been investigated. The local Nusselt number is, $Nu_L = -\mu\left(\frac{\partial \theta}{\partial Y}\right)_{Y=0}$.

## NUMERICAL TECHNIQUE

In this section, the governing second order coupled dimensionless partial differential equations with initial and boundary conditions have been solved. The explicit finite difference method has been used to solve equations (1) to (4) subject to the boundary conditions. The region of the flow is divided into a grid or mesh of lines parallel to $X$ and $Y$-axes where $X$-axis is taken along the plate and $Y$ axis is normal to the plate.

It is assumed that the plate of height $X_{\max}(=60)$ i.e. $X$ varies from 0 to 60 and regard $Y_{\max}(=20)$ as corresponding to $Y \to \infty$ i.e. $Y$ varies from 0 to 20. The number of grid space in $X$ and $Y$ directions are $m = 60$ and $n = 60$. Hence the constant mesh sizes along $X$ and $Y$ axes becomes $\Delta X = 1.0 (0 \le x \le 60)$ and $\Delta Y = 0.33 (0 \le y \le 20)$ with smaller time-step $\Delta \tau = 0.01$.

Let, $U'$, $W'$ and $\theta'$ denote the values of $U$, $W$ and $\theta$ at the end of a time-step respectively. Using the explicit finite difference approximation, the appropriate set of finite difference equations are obtained as follows:

$$\frac{U_{i,j} - U_{i-1,j}}{\Delta X} + \frac{V_{i,j} - V_{i,j-1}}{\Delta Y} = 0 \tag{5}$$

$$\frac{U'_{i,j} - U_{i,j}}{\Delta \tau} + U_{i,j}\frac{U_{i,j} - U_{i-1,j}}{\Delta X} + V_{i,j}\frac{U_{i,j} - U_{i,j-1}}{\Delta Y} = \frac{U_{i,j+1} - 2U_{i,j} + U_{i,j-1}}{(\Delta Y)^2} + Ze^{-Y} + 2RW_{i,j} \tag{6}$$

$$\frac{W'_{i,j} - W_{i,j}}{\Delta \tau} + U_{i,j}\frac{W_{i,j} - W_{i-1,j}}{\Delta X} + V_{i,j}\frac{W_{i,j} - W_{i,j-1}}{\Delta Y} = \frac{W_{i,j+1} - 2W_{i,j} + W_{i,j-1}}{(\Delta Y)^2} + Ze^{-Y} - 2RU_{i,j} \tag{7}$$

$$\frac{\theta'_{i,j} - \theta_{i,j}}{\Delta \tau} + U_{i,j}\frac{\theta_{i,j} - \theta_{i-1,j}}{\Delta X} + V_{i,j}\frac{\theta_{i,j} - \theta_{i,j-1}}{\Delta Y} = \left(\frac{1}{P_r} + \frac{4}{3}R_d\right)\frac{\theta_{i,j+1} - 2\theta_{i,j} + \theta_{i,j-1}}{(\Delta Y)^2} + E_c\left[\left(\frac{U_{i,j} - U_{i,j-1}}{\Delta Y}\right)^2 + \left(\frac{W_{i,j} - W_{i,j-1}}{\Delta Y}\right)^2\right] \tag{8}$$

And the initial and boundary conditions with the finite difference scheme are given as follows:

$t > 0$, $U_{i,L} = 0$, $W_{i,L} = 0$, $\theta_{i,L} = 0$ at $L = 0$ and $U_{i,L} = 0$, $W_{i,L} = 0$, $\theta_{i,L} = 1$ at $L \to \infty$

Here the subscripts $i$ and $j$ designate the grid points with $X$ and $Y$ coordinates respectively.

## STABILITY AND CONVERGENCE ANALYSIS

Since an explicit procedure is being used, the analysis will remain incomplete unless the stability and convergence of the finite difference scheme are discussed. For the constant mesh sizes, the stability criteria finally can be written as: $\frac{U\Delta\tau}{\Delta X} - \frac{|V|\Delta\tau}{\Delta Y} + \frac{2\Delta\tau}{P_r(\Delta Y)^2} + \frac{4}{3}R_d\frac{2\Delta\tau}{(\Delta Y)^2} \le 1$

Using $\Delta Y = 0.33$, $\Delta \tau = 0.01$ and the initial condition, the above equations gives $P_r \geq -0.13$, $R_d \leq 10$, $E_c = 0.10$ and $0 < R \leq 0.1$.

## RESULTS AND DISCUSSION

In order to investigate the physical condition of the developed mathematical model, the steady-state numerical results have been computed for the non-dimensional primary velocity $(U)$, secondary velocity $(W)$ and temperature $(\theta)$ within the boundary layer. The effect of modified Hartmann number $(Z)$ and Rotational parameter $(R)$ on velocity and temperature distributions as well as local shear stresses and local Nusselt number are discussed, which are shown in Figs. 3 and 4. For brevity, the effect of other parameters such as Prandtl number $(P_r)$, Eckert number $(E_c)$ and Radiation parameter $(R_d)$ are not shown. In Figs.3(c) and 4(c), the enlargement of temperature distribution is illustrated.

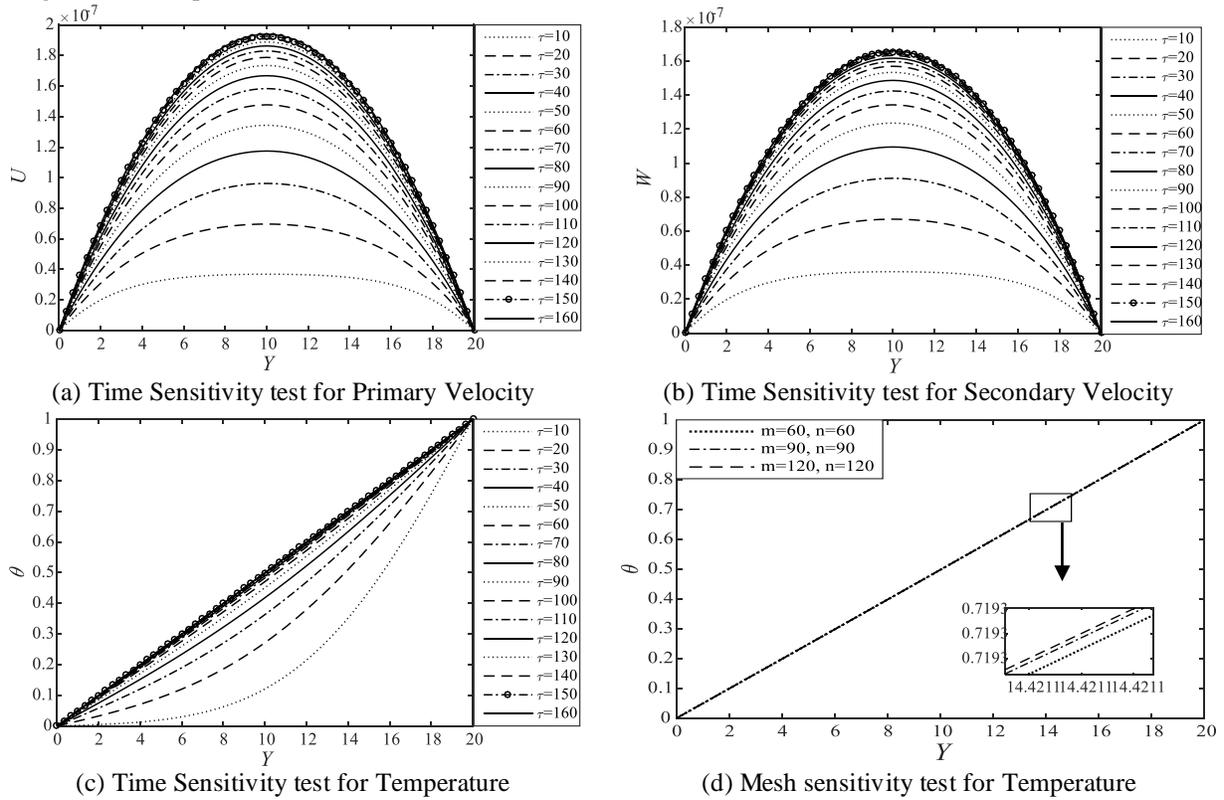

(a) Time Sensitivity test for Primary Velocity
(b) Time Sensitivity test for Secondary Velocity
(c) Time Sensitivity test for Temperature
(d) Mesh sensitivity test for Temperature

**Figure 2.** Sensitivity test, where $Z = 2.50$, $P_r = 0.71$, $E_c = 0.10$, $R = 0.001$ and $R_d = 2.00$

**Time sensitivity test:** To obtain the above-mentioned steady-state solution of the developed mathematical model, the computations for the primary velocity, secondary velocity, and temperature profiles have been continued for different dimensionless time $(\tau)$. It is observed that the result of computations for different profiles, however, shows little changes after $\tau = 120$ and shows negligible changes after $\tau = 150$. Thus, the solutions of all variables for $\tau = 150$ are taken essentially as the steady-state solutions. Fig. 2(a,b,c) shows that the primary velocity, secondary velocity, and temperature profiles reach their steady state monotonically. It also should be mentioned that the temperature profile reaches the steady state faster than both the primary and secondary velocity profiles. Also, the primary velocity profile reaches the steady state smoothly than the secondary velocity profile.

**Mesh sensitivity test:** To obtain the appropriate mesh space for $m$ and $n$, the computations have been carried out for three different mesh spaces such as $(m,n) = (60, 60), (90, 90), (120, 120)$ as shown in Fig. 2(d). The curves are smooth for all mesh spaces and shows a negligible changes among the curves. Thus the mesh size $(m,n) = (60, 60)$ can be chosen as the appropriate mesh space.

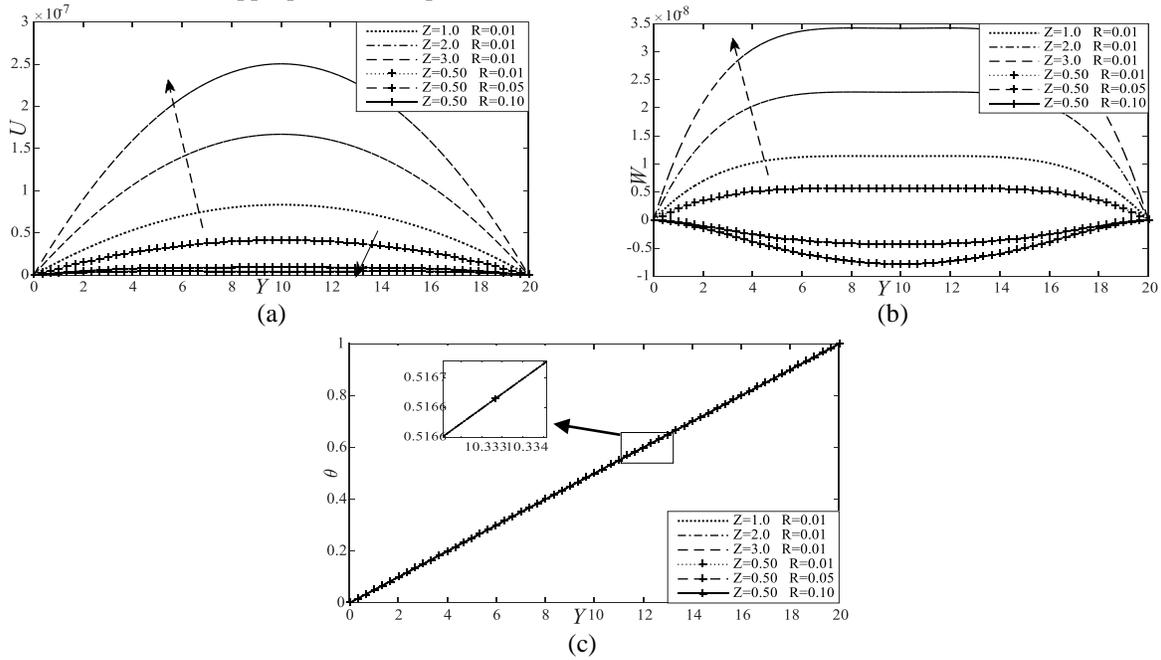

**Figure 3.** Effect of modified Hartmann number $(Z)$ and Rotational parameter $(R)$ on (a) Primary velocity; (b) Secondary velocity and (c) Temperature distributions; where $E_c = 0.10$, $P_r = 0.50$ and $R_d = 2.00$ at time $\tau = 150$ (Steady State)

Fig. 3 shows that the primary velocity and secondary velocity both increases with the increase of $(Z)$ while the primary velocity decreases with the increase of $(R)$ and the secondary velocity fluctuates with the increase of $(R)$. Furthermore, temperature distributions have negligible effect with the increase of $(Z)$ and $(R)$ both.

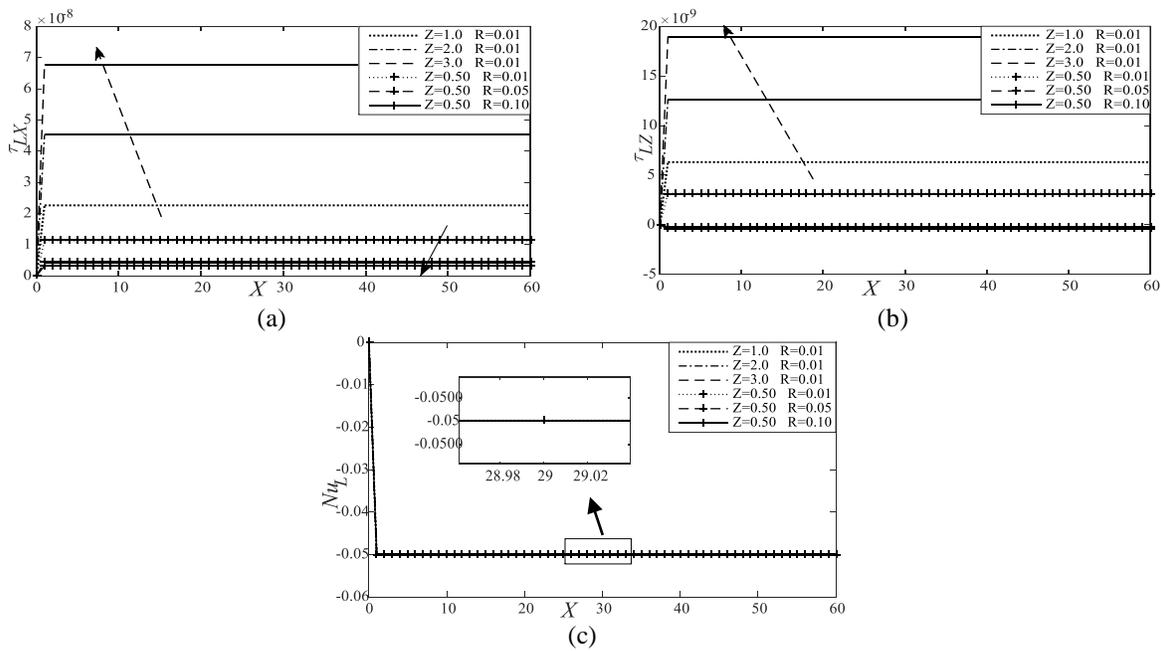

**Figure 4.** Effect of modified Hartmann number $(Z)$ and Rotational parameter $(R)$ on (a) local primary shear stress; (b) local secondary shear stress and (c) local Nusselt number; where $E_c = 0.10$, $P_r = 0.50$ and $R_d = 2.00$ at time $\tau = 150$ (Steady State)

Fig. 4 shows that the local primary shear stress and local secondary shear stress both increases with the increase of $(Z)$ while the local primary shear stress decreases with the increase of $(R)$ and the local secondary shear stress fluctuates with the increase of $(R)$. Furthermore, the local Nusselt number has negligible effects with the increase of $(Z)$ and $(R)$.

## CONCLUSIONS

The explicit finite difference solution for unsteady viscous incompressible fluid flow along the Riga plate with the influence of magnetic force in a rotating system has been investigated. The numerical technique has found to be converged for $P_r \geq -0.13$, $R_d \leq 10$, $E_c = 0.10$ and $0 < R \leq 0.1$. The results are discussed for different values of important parameters as modified Hartmann number $(Z)$ and Rotational parameter $(R)$. For brevity, the effect of other parameters such as Prandtl number $(P_r)$, Eckert number $(E_c)$ and Radiation parameter $(R_d)$ are not shown. Based on the results and discussion, some important findings are mentioned as follows:

1. The primary velocity and local primary shear stress both are increased with the increase of $(Z)$ while both are decreased with the increase of $(R)$.
2. The secondary velocity and local secondary shear stress both are increased with the increase of $(Z)$ while both have fluctuated with the increase of $(R)$.
3. The temperature profiles and the local Nusselt number both have changed negligibly with the increase of $(Z)$ and $(R)$ both.
4. The temperature profile reaches the steady state faster than both the primary and secondary velocity profiles.

## ACKNOWLEDGMENTS

It is partially possible to complete the research work with the financial support of NST Fellowship under the Ministry of Science and Technology, Government of the People's Republic of Bangladesh.